\documentclass[usenatbib]{mn2e}
\usepackage{graphicx}
\usepackage{ifthen}

\def\ltsima{$\; \buildrel < \over \sim \;$}
\def\simlt{\lower.5ex\hbox{\ltsima}}
\def\gtsima{$\; \buildrel > \over \sim \;$}
\def\simgt{\lower.5ex\hbox{\gtsima}}
\def\kms{{\rm\,km\,s^{-1}}}

\def\kpc{{\rm\,kpc}}

\def\msun{{\rm\,M_\odot}}

\def\pc{{\rm\,pc}}

\def\CompactFigs{0}
\def\UseFigs{1}

\def\s{\ifmmode \widetilde \else \~\fi}
\def\={\overline}

\def\spose#1{\hbox to 0pt{#1\hss}}

\def\lta{\mathrel{\spose{\lower 3pt\hbox{$\mathchar"218$}}
     \raise 2.0pt\hbox{$\mathchar"13C$}}}
\def\gta{\mathrel{\spose{\lower 3pt\hbox{$\mathchar"218$}}
     \raise 2.0pt\hbox{$\mathchar"13E$}}}
\def\Dt{\spose{\raise 1.5ex\hbox{\hskip3pt$\mathchar"201$}}}    
\def\dt{\spose{\raise 1.0ex\hbox{\hskip2pt$\mathchar"201$}}}    

\def\dotsfill{\leaders\hbox to 1em{\hss.\hss}\hfill}

\def\Gyr{{\rm\,Gyr}}

\loadboldmathitalic 
\title[Uncovering CDM halo substructure]{Uncovering CDM halo substructure with tidal streams}
\author[R. A. Ibata, G. F. Lewis \& M. J. Irwin]
{R. A. Ibata$^{1}$, G. F. Lewis$^{2}$ \& M. J. Irwin$^{3}$ \\
$^{1}$
Observatoire de Strasbourg, 11, rue de l'Universit\'e, F-67000, Strasbourg, 
France\\
$^{2}$
Anglo-Australian Observatory, P.O. Box 296, Epping, NSW 1710, Australia\\
$^{3}$
Institute of Astronomy, Madingley Road, Cambridge, CB3 0HA, U.K.}

\date{\today}
\begin{document} 
\maketitle 
\begin{abstract}
Models  for the  formation and  growth of  structure in  a cold  dark matter
dominated  universe predict  that  galaxy halos  should contain  significant
substructure. Studies  of the Milky Way,  however, have yet  to identify the
expected few  hundred sub-halos with  masses greater than  about $10^6\msun$.
Here  we propose  a  test for  the presence  of  sub-halos in  the halos  of
galaxies.  We show that the structure of the tidal tails of ancient globular
clusters is very sensitive to  heating by repeated close encounters with the
massive dark sub-halos.   We discuss the detection of such  an effect in the
context of the next generation of astrometric missions, and conclude that it
should be easily detectable with the  GAIA dataset.  The finding of a single
extended  cold  stellar  stream   from  a  globular  cluster  would  support
alternative theories,  such as self-interacting dark matter,  that give rise
to smoother halos.
\end{abstract}

\section{Introduction}

A generic prediction of Cold Dark Matter (CDM) cosmology is the existence of
a significant amount of substructure in gravitationally collapsed structures
such as galaxy clusters, galaxy halos and even dwarf galaxies \citep{klypin,
moore99b, moore01}.  On the  scale of large  spiral galaxies like  the Milky
Way,  some 500  dense  clumps are  expected  to orbit  in  the Halo.   These
structures  have very  dense  cores, with  a  very steep  radial profile  of
universal  shape  (increasing  as  $r^{-1}$  or $r^{-1.5}$  in  the  central
regions,  \citealt{navarro,  moore99a}),   which  renders  them  essentially
impervious to Galactic tides.

The current ``dark  matter crisis'' stems from the  difference between these
predictions  and observations of  low-mass galaxies.   The first  problem is
that rotation curves of gas-rich low-mass disk galaxies are not as peaked as
required by CDM.  \citet{vandenbosch} suggested that this disagreement could
be  due to  an  error  in the  analysis  of HI  observations,  in which  the
beam-smearing effect of extant HI  radial velocity curves had not been taken
into account.   However, this appears  not to be  the case, as  recent (high
spatial  resolution)  $H_\alpha$  rotation  curves have  shown  \citep{cote,
deblok, marchesini}.

A second fundamental problem also  exists: the mass of the CDM substructures
in a  galaxy-sized halo is similar  to that inferred  for galaxy satellites,
yet the  number of CDM substructures  in a galaxy-sized halo  is much larger
than the  number of observed dwarf galaxies\citep{klypin,  moore99b}.  It is,
of   course,  possible   that  these   galaxies  have   a   quite  different
non-cosmological  origin (perhaps  due to  tidal interactions  in  the early
universe).  In either  case the problem remains: where  is the population of
500  CDM substructures  orbiting the  Milky  Way and  other large  galaxies?
Recently, several  studies have investigated the effect  of reioninzation on
the early evolution of  small structures \citep{bullock, somerville, tully}.
This  re-ionizing  radiation sets  in  at  a  redshift $\sim  6$--$10$,  and
originates from either the first stars or quasars.  Indeed, the first direct
evidence  for   reionization  (at  $z\sim   6$)  has  recently   been  found
\citep{becker}.  The intergalactic and  galactic mediums are ionized by this
radiation, and any gas that is not  in deep potential wells is lost from the
protogalaxies.  The limiting mass for maintaining a gas component appears to
correspond to a circular velocity  of $\sim 30\kms$, well above the circular
velocity of  all the dwarf spheroidal  galaxies known.  (However,  it may be
possible for low-mass systems to  maintain a gas-fraction if the gas managed
to cool  to dense  molecular form before  the epoch of  reionization).  This
``squelching'' of star-formation in  low-mass galaxies neatly solves the CDM
satellite over-production problem.

A potential  candidate for  the missing dwarf  systems that are  supposed to
inhabit the  Galactic halo is the  large population of  high velocity clouds
that pepper the sky (see Wakker 2001 for a recent review). With little or no
stellar  content, these  are seen  to possess  velocities  incompatible with
Galactic  rotation  models,  although  \citet{blitz}  have  shown  that  the
kinematic  characteristics of the  velocities of  the clouds  are compatible
with  them being  members of  an accreting  population distributed  within a
megaparsec of the  Milky Way.  At these distances,  the kinematic signatures
of the clouds imply a mass-to-light ratio of 10-50 \cite{braun}.  Currently,
their distances are  extremely difficult to determine, with  Complex A being
the only  system to  which the distance,  of between  4 and 10kpc,  has been
measured  \citep{woerden}.   With the  promise  of  accurate distances  from
$H_\alpha$ \citep{bland99}, a full appraisal  of the nature of high velocity
clouds and their relation to the formation of the Milky Way can be made; but
a  preliminary  analysis suggests  that  the  clouds  are scattered  in  the
Galactic halo \citep{bland01}, and are not of cosmological origin.

Nevertheless, CDM theory  predicts that a large population  of completely or
almost  completely dark  ``galaxies'' should  be present  in the  Milky Way.
Here we propose an observational test of this scenario. In the early galaxy,
many  globular clusters  are  expected  to have  surrounded  the Milky  Way.
Galactic tidal forces have been destroying these objects \citep{gnedin}, and
of the initial population of a  few hundred objects, only $\sim 100$ remain,
and many of these show signs of tidal disruption \citep{leon, odenkirchen}.

In a  static potential,  as a  low-mass stellar system,  such as  a globular
cluster, loses progressively more and more mass to tidal forces, it develops
long tidal tails that closely follow the orbit of the globular cluster.  (In
the  limit of  a zero  mass  system, the  tidal stream  exactly follows  the
progenitor orbit).   Here we explore how  the presence of  the population of
dark matter  clumps predicted by CDM  alters the phase-space  structure of a
globular  cluster  tidal stream.   For  the  special  case when  the  global
Galactic potential is nearly spherical,  this corresponds to a broadening of
the stream from a thin great-circle stream into a wide band on the sky. In a
companion  paper  (\citealt{ibata02};  Paper~2),  we use  these  results  to
motivate the search for tidal streams in the 2~Micron All Sky Survey (2MASS)
dataset.

Other  ways  to probe  the  dark  matter  substructure have  been  proposed.
Recently, \citet{metcalf} and  \citet{chiba2} suggested that substructure in
external  galaxies may be  detected through  their gravitational  lensing of
background  quasars, as  a  lumpy  halo will  produce  image and  brightness
configurations different  to a smooth  matter halo.  While such  an approach
may reveal any  missing sub-halos, the method is subject  to the vagaries of
gravitational  lens modeling.  During  the final  preparation of  this work,
\citet{mayer}  presented  a  study  investigating the  signatures  that  are
imprinted  by  different  dark  matter  models on  tidal  streams  of  dwarf
spheroidal galaxies.   They find  that the structure  and kinematics  of the
outer regions of dwarf galaxies  can also allow one to differentiate between
dark matter models.

\section{Numerical simulations}

Our aim is to quantify the difference  in the structure of the tidal tail of
an ancient globular cluster (or even an ancient globular cluster remnant) if
the Galactic halo  has a smooth mass distribution compared  to the case when
the Halo also contains significant substructure. We have modelled the smooth
components of the Galaxy as  fixed potentials, using the Galactic mass model
of \citet{dehnen},  which contains a disk, thick  disk, interstellar medium,
bulge,  spheroid and  halo components.   The  parameters of  this model  are
detailed in \citet{ibata}.  The halo  was taken to have a mass normalization
that gives a total model circular velocity of $v_c=210\kms$ at $50\kpc$. The
mass flattening of the potential was left as a free parameter $q_m$.

We adapted the fast parallel code PKDGRAV to include the forces due to these
mass distributions  by including a multipole expansion  code kindly supplied
by  W. Dehnen.   In all  the  following simulations  we maintained  accurate
forces by using an opening angle of $0.75$ and expanding the cell moments to
hexadecapole order.   Two body relaxation  was suppressed by using  a spline
softening  of 10  pc, such  that  the inter-particle  forces are  completely
Newtonian at 20 pc.  A variable  time-step scheme is used based on the local
acceleration, $\Delta  t < \eta  \sqrt{|\Phi|}/a$, and density, $\Delta  t <
\eta/\sqrt{G   \rho}$,   with  the   accuracy   parameter   $\eta  =   0.03$
\citep{quinn}.  Typically  we used  100000 base steps  to integrate  12 Gyr.
With the variable  time-steps, this was equivalent to  taking $5\times 10^7$
time-steps.

The initial  globular cluster was  modelled with a King  model \citep{king},
populated with $10^4$ particles. The  total mass of the globular cluster was
taken to be  $10^6\msun$, the model had concentration  parameter $c=1.0$ and
central  velocity dispersion  $\sigma=4.1\kms$, yielding  a tidal  radius of
$r_t=300\pc$. This model is of approximately the same size and mass as Omega
Centauri, but of  lower concentration (by a factor of  $1.7$).  Note that we
are  not aiming  at presenting  an  accurate model  of the  disruption of  a
globular  cluster ---  we do  not have  sufficient resolution  for  that ---
instead, here we seek to study  the kinematic behaviour of the unbound tidal
streams that emerge from disrupting low-mass systems.

The initial position  and systemic velocity of the  King model were randomly
chosen from the halo component of  a galaxy model \citep{boily} that gives a
similar rotation curve to the adopted Dehnen \& Binney mass model.

Integrating the globular cluster models for $10\Gyr$ in the smooth potential
gives rise to very narrow tidal tails.  An example of one of these models is
shown in Figure~1, which displays the position, heliocentric radial velocity
and distance  distribution of  the stream particles.  At its  narrowest, the
width of this stream is $\sim  100\pc$, similar in width to the tidal radius
of the initial  King model, while at its widest, the  stream is $\sim 3\kpc$
wide.  The shape of the  halo mass distribution for the simulation displayed
in Figure~1 is spherical, $q_m=1$,  which is motivated by recent analyses of
the outer halo of the Milky Way \citep{ibata,chiba}.

For each one of these models, we also investigate the effect of the presence
of a  large number of moving  substructures in the halo.   However, we first
need a realistic way of modeling these substructures.  According to Navarro,
Frenk \& White (1997) (hereafter  NFW), and other authors, galaxy halos have
a ``universal''  density profile $\rho(r) \propto  1/(r(1+r)^2)$, which fits
all  scales currently probed  by the  numerical cosmology  simulations.  The
potential  corresponding to this  density profile  is very  simple: $\phi(r)
\propto  1-\ln(1+r)/r$,  and the  forces  due  to  this potential  would  be
straight-forward to add in to our N-body integrator. Unfortunately, however,
this density profile  is not well-behaved either at small  or at large radii
(the total mass diverges); to fix  these problems, the force near the center
of the sub-halo would need to be  softened, and the density would have to be
made to  fall off  to zero  at some large  radius (the  tidal radius  of the
substructure would be a good  choice).  A simpler solution is to approximate
the NFW  profiles with a  softened point-mass potential.   Figure~2 compares
the  radial acceleration due  to an  NFW potential  (dotted line),  a Kepler
potential (dashed line),  and a softened Kepler potential  (solid line) with
spline softening of $3.41$ times the  scale radius $r_s$ of the NFW profile.
Note that,  at all radii, this softened  point-mass potential underestimates
the forces compared to the NFW potential.

The substructures are added into the Galactic halo as softened point masses,
with   a  distribution   of  sub-halo   masses  that   follows   a  relation
$\log_{10}(N_c)=3  (  1  -  4  v_c/V_{Global} )$  (our  parameterization  of
Figure~2 of  \citealt{moore01}).  At the low-mass end  of this distribution,
there are 435 sub-halos of  circular velocity $v_c>0.03 V_{Global}$ ($M \sim
10^7 \msun$).  The halo mass  fraction in this lumpy component is relatively
small, as it accounts  for less than $10$\% of the mass  of the smooth halo.
The  initial positions  and  velocities of  the  particles representing  the
sub-halos  are also  chosen from  the  halo component  of the  \citet{boily}
Galaxy  model, but  with a  pericenter cutoff  of $10\kpc$,  since dynamical
friction  will  likely make  the  orbits  of  dark satellites  with  smaller
pericenter radii decay quickly.

The effect  of these numerous dark  matter clumps is marked:  in Figure~3 we
show the  position, velocity and  distance distribution of the  tidal stream
particles. In this  simulation, only 1\% of  the mass of the halo  is not in
the smooth component, yet the tidal stream has been substantially heated and
is substantially wider than in the smooth halo (the stream is so fluffy that
it is difficult to measure a width).

A better way to look at the  simulations is in the space of the integrals of
the motion.   In the axisymmetric potentials considered  here, the integrals
are the  total energy  per unit mass  $E$ and  the $z$ component  of angular
momentum  per unit  mass $L_z$.   It  is also  useful to  inspect the  total
angular momentum per unit mass, $L$, which is an approximate integral of the
motion in  the relatively spherical  potentials we consider  here.  Figure~4
shows the relation between these  three quantities.  The top row of Figure~4
is derived from  the simulation of the globular cluster  model in the smooth
spherical  halo whose  structure  on  the sky  was  previously displayed  in
Figure~1. After $12\Gyr$ the particles  have remained localised in the space
of the  integrals of the  motion. The bottom  panel shows the  same globular
cluster  model on  the same  orbit,  but simulated  in the  halo model  that
contains the additional NFW sub-halos (corresponding to the simulation shown
in  Figure~3). In  this lumpier  potential, the  globular cluster  stream is
comparatively much  more dispersed, especially  in the $L_z$  parameter. The
r.m.s.   dispersion in  $L_z$ is  a factor  of 5  larger in  the  lumpy halo
simulation.

Inspecting  the  spatial  distribution   of  particles  can  be  useful  for
differentiating between a  smooth and lumpy halo when  the mass distribution
is  spherical  (or  close  to  spherical). However,  in  a  flattened  halo,
differential precession mimics  the effect of heating by  CDM clumps, and it
becomes  impossible  to differentiate  between  the  two  causes of  spatial
dispersion of the  stream. It is only when viewing the  space of the integrals
of the  motion that the effect  of precession decouples  from the stochastic
heating by the  CDM clumps.  We have repeated the  same experiments as above
in  a  flattened halo,  with  $q_m=0.7$,  a  flattening value  predicted  by
numerical  cosmology \citep{katz91a,  dubinski91, warren,  katz91b, summers,
dubinski94}.   The sub-halos are  chosen from  a halo  distribution function
which is also  flattened to $q_m=0.7$.  Figure~5 shows  the relation between
the integrals of the motion $E$  and $L_z$ and the approximate integral $L$.
Again,  the dispersing  effect of  the sub-halos  is very  strong,  with the
r.m.s.  dispersion in $L_z$ increasing by  a factor of 9 with respect to the
smooth halo case.

\subsection{Suite of simulations}

To obtain a statistical understanding of the change in $L_z$, we undertook a
series of  simulations of the  globular cluster model with  initial position
and velocity  randomly-drawn from  the spherical halo  distribution function
model mentioned above.  Ten simulations  were undertaken in the Galaxy model
with the  smooth spherical halo. A  further ten simulations  probed the same
models with the addition of the NFW sub-halos, for a total of 20 simulations
with  $q_m=1.0$.  To  probe the  situation with  a more  flattened  halo, we
undertook a further 20 simulations with $q_m=0.7$.

The result of  these experiments is shown in Figure~6,  where we display the
r.m.s. dispersion in $L_z$ for simulations in a smooth halo (filled circles)
and in in the presence of  halo substructures (star symbols).  The top panel
shows the  case in a spherical  halo ($q_m=1.0$), while the  bottom panel is
for the flattened halo ($q_m=0.7$) simulations.

Both  in a  spherical and  in  a flattened  halo, the  presence of  CDM-like
substructures  increases substantially,  on average,  the RMS  dispersion in
$L_z$ of  the stream stars. On some  orbits (such as that  of our simulation
\#9,  the  effect can  be  small,  however.   Our simulations  suggest  that
globular cluster  streams with a  very dispersed specific  angular momentum,
say $\sigma L_z  > 200 \kpc \kms \msun^{-1}$ require  that the halo possesses
significant  substructure.   Smaller values,  $\sigma  L_z  <  50 \kpc  \kms
\msun^{-1}$ say, would require a smooth halo.

\section{Observational considerations}

We  next consider  whether it  will be  possible to  identify  a $10^6\msun$
stream  orbiting the  Halo  at  distances up  to  $100\kpc$ ($m-M=20$)  with
photometry, radial velocity, parallax and proper motion data from the future
astrometric mission GAIA \citep{perryman}.   The great advantage of the GAIA
dataset  is  that  gives  access  to  the  full  6-dimensional  phase  space
information  for   many  stars,  and  at   least  4-dimensional  phase-space
information (for  those stars that are  too far to  have measured parallaxes
and  radial velocities).  In  addition, the  photometric information  (in 15
bands)  can give  age  and/or metallicity  estimates,  providing a  powerful
discriminant between stellar populations.

Developing a  full model of the  Galaxy as viewed  by GAIA is not  a trivial
task and it is beyond the scope  of this paper. Here we simply make an order
of magnitude estimate of the  contrast of plausible globular cluster streams
over the background by estimating the relative phase space density of stream
stars over  the large-scale Galactic  spheroid (or stellar  halo) component.
We estimate the phase-space size of the stream as follows. The $\sim 10\kms$
radial velocity  uncertainty of GAIA  exceeds the intrinsic  radial velocity
dispersion of  the stream,  and is  about 10\% of  the halo  radial velocity
dispersion.  The proper motion  dispersion at $100\kpc$ corresponds to about
$\sim  10\kms$  or 10\%  of  the  halo  dispersion (two  coordinates).   The
positional width  of the  streams are  $\simlt 3\kpc$, or  about 1\%  of the
width  of  the  halo. With  the  full  15  photometric bands,  the  distance
uncertainty of photometric  parallax measurements is likely to  be $< 10$\%.
Thus the phase-space volume occupied by the stream is likely to be about one
millionth of that  of the halo.  Since  the stellar halo has a  mass of only
$\sim 10^9\msun$, the contamination will amount to approximately one star in
a thousand.  With a limiting magnitude of $V \sim 21$, the GAIA measurements
will easily reach down to the horizontal branch at $100\kpc$. Placing all of
the stream at a  distance of $d = 100\kpc$, we expect  $\sim 1600$ stars for
$V  > 21$  per  $10^6 \msun$  of  progenitor mass  (assuming the  luminosity
function of  the Draco by Odenkirchen et  al.  2001, corrected to  a mass to
light  ratio of  $M/L=3$).  For  radial velocity  measurements  the limiting
magnitude is $V\sim  17$, and the number of available  stream stars from the
$10^6\msun$ population drops  to $\sim 180$. The detection  of these streams
will be  straightforward with  GAIA, and the  increase in the  dispersion of
$L_z$ due to any halo substructure will be easily resolved.

With results expected towards the end  of the second decade of this century,
the  GAIA  mission  is not  for  the  impatient.   So  it is  worthwhile  to
investigate  whether  any   presently  available  datasets  already  contain
sufficient  information   to  uncover  the  predicted   Galactic  halo  dark
substructures. To  this end we undertook  an analysis of  the 2MASS dataset,
which provides  positions and infrared  CCD photometry for  sources brighter
than  $K=14.3$.  This  gives access  to  a relatively  restricted sample  of
Galactic halo M-giants, with no  information on their kinematics, and only a
very poor constraint on their distances.  The search for the halo streams in
the 2MASS  dataset is  presented in Paper~2;  the analysis reveals  a single
very strong  stream, that due to  the ongoing disruption  of the Sagittarius
dwarf galaxy.

Is it possible  to use the Sagittarius stream to  constrain the lumpiness of
the halo?   To investigate this  issue we ran  two simulations, in  the same
manner as above, one in a smooth  halo, and the other with NFW clumps, using
as  an   initial  model   for  the  Sagittarius   dwarf  the  model   D2  of
\citet{ibata}. Figure~7 shows the result  of this experiment.  This model is
much  more massive  ($M=5\times10^8\msun$) than  the globular  cluster model
simulated above,  and is initially  much more extended (half-mass  radius of
$0.9\kpc$),  so it  is naturally  widely dispersed  in $E$,  $L_z$  and $L$.
Though the stream  particles are heated and dispersed  by the sub-halos, the
effect is subtle, and will be  very difficult to disentangle from the normal
dynamical evolution of the dwarf  galaxy.  So low mass streams from globular
clusters need to be found to act as probes of the dark matter.

\section{Conclusions}

We have  shown that stellar  streams of low  mass systems such  as disrupted
globular  clusters can be  used to  constrain the  substructure of  the dark
halo.   If  current cosmological  simulations  correctly represent  reality,
ancient  stellar  streams from  globular  clusters  should be  substantially
dispersed over the  sky, even if on average, the  Galactic halo potential is
close to spherical.  This effect  is readily noticeable, as the streams will
have  a large  dispersion in  the z-component  of specific  angular momentum
$L_z$. For  the particular halo studied  in this contribution  $\sigma L_z >
200 \kpc  \kms \msun^{-1}$ implies  the presence of substantial  dark matter
substructure.

We  have  simplified  the  analysis  in two  major  ways.   Highly  softened
point-masses were  used to model the  halo substructures, which  leads to an
underestimate of  the increased dispersion in  $L_z$ in the  presence of the
NFW  clumps.  Our  other major  assumption, that  the Galactic  potential is
axisymmetric and  does not evolve with  time, will likely  have the opposite
effect, namely  to underestimate  the dispersion in  the absence of  the NFW
clumps.  Our study  should be refined once the  Galactic potential is better
constrained (with data from the GAIA and SIM missions).

This work has  shown that an all-sky survey  with high precision astrometry,
such as will be obtained with the  GAIA satellite, we will easily be able to
detect  globular cluster  streams, and  measure their  kinematic properties,
especially the dispersion in $L_z$.

The heating due  to the numerous encounters with  the dark substructure must
also affect the computed destruction  rate of globular clusters and low-mass
dwarf galaxies  alike. It is  possible that the  lower limit to the  mass of
dSph galaxies $\sim 10^7\msun$ may be set by these disruptive encounters.

Discovering and  determining the thickness  of ancient stellar  streams will
allow us to probe the invisible  dark halo structures predicted by Cold Dark
Matter cosmology. Finding a single  narrow stream of width comparable to the
tidal radius of a globular  cluster would stand in strong contradiction with
this theory. Data from the  forthcoming GAIA mission planned by the European
Space Agency will provide the opportunity to undertake this study.

\newcommand{\mnras}{MNRAS}
\newcommand{\nat}{Nature}
\newcommand{\araa}{ARAA}
\newcommand{\aj}{AJ}
\newcommand{\apj}{ApJ}
\newcommand{\apjl}{ApJ}
\newcommand{\apjs}{ApJSupp}
\newcommand{\aap}{A\&A}
\newcommand{\aaps}{A\&ASupp}
\newcommand{\pasp}{PASP}


\begin{figure}
\ifthenelse{\UseFigs=1}{
\ifthenelse{\CompactFigs=0}
{\includegraphics[width=\hsize]{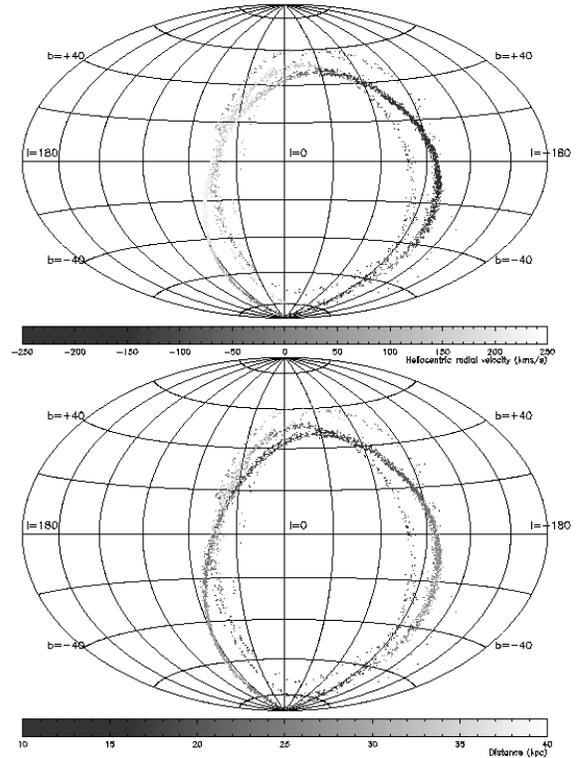}}
{\includegraphics[width=\hsize]{GAIAcdm.fig01c.ps}}}{}
\caption{The  sky  distribution (in  Galactic  coordinates)  of  one of  the
globular  cluster models after  $12\Gyr$.  The  Galactic potential  has been
modelled as  a sum  of a  disk, thick disk,  interstellar medium,  bulge and
spheroid components, plus a spherical  dark halo.  In this simulation all of
these Galactic components have potentials that are smooth, axisymmetric, and
static.   The upper  diagram shows  the  particle velocities  and the  lower
diagram shows the particle distances.}
\end{figure}

\begin{figure}
\ifthenelse{\UseFigs=1}{
\includegraphics[width=\hsize]{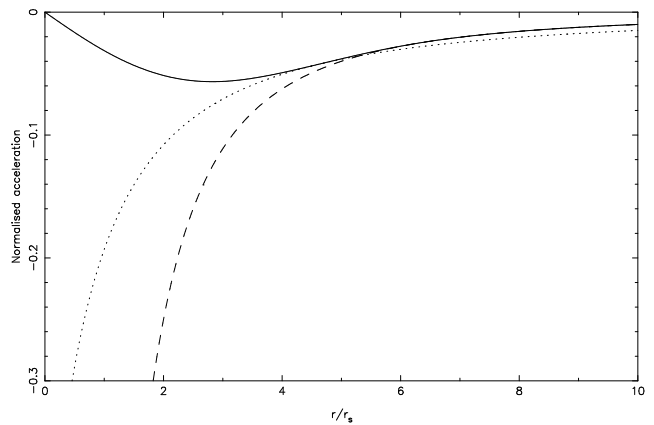}}{}
\caption{The  acceleration as a  function of  radius due  to an  NFW halo
(dotted line)  a point-mass (dashed  line) and a spline-softened  point mass
with $r_{soft} = 3.41 r_s$. By modeling the NFW sub-halos as softened point
masses, we underestimate  the forces (and hence the  heating) they impart on
other particles in the simulation.}
\end{figure}

\begin{figure}
\ifthenelse{\UseFigs=1}{
\ifthenelse{\CompactFigs=0}
{\includegraphics[width=\hsize]{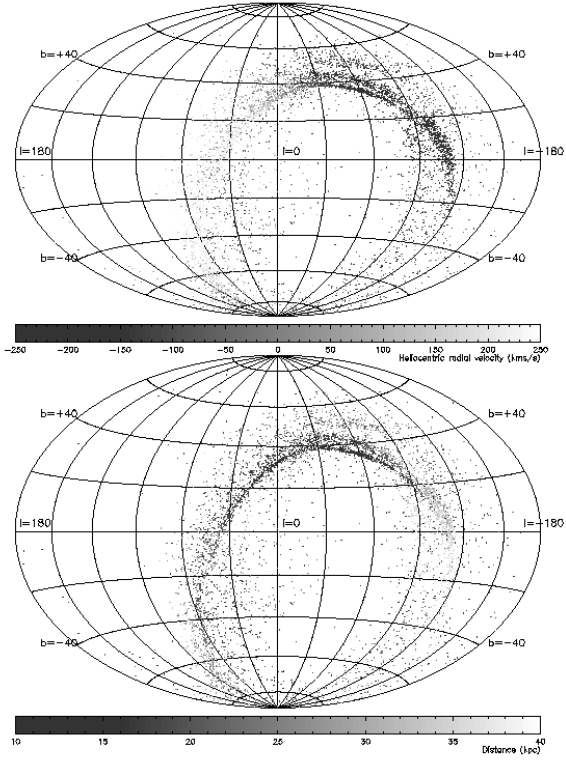}}
{\includegraphics[width=\hsize]{GAIAcdm.fig03c.ps}}}{}
\caption{As Figure~1, but this time  the globular cluster has been simulated
in a Galactic potential replacing a small mass fraction of the halo with 435
moving NFW sub-halos.}
\end{figure}

\begin{figure}
\ifthenelse{\UseFigs=1}{
\includegraphics[angle=270,width=\hsize]{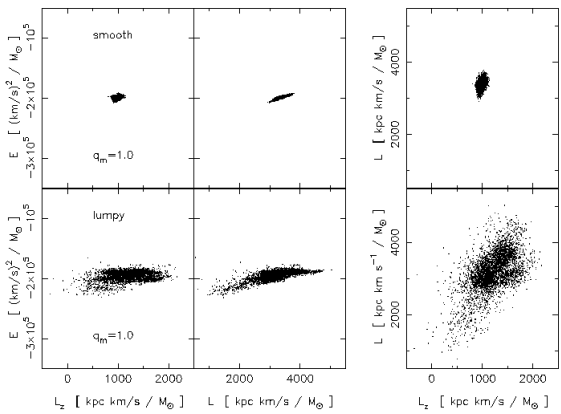}}{}
\caption{The relation between the integrals of motion for the simulation
displayed in Figure~1 (top row) and the simulation displayed in Figure~3
(bottom row). The addition of a small fraction by mass in lumpy sub-halos
has completely smeared out the stream.}
\end{figure}

\begin{figure}
\ifthenelse{\UseFigs=1}{
\includegraphics[angle=270,width=\hsize]{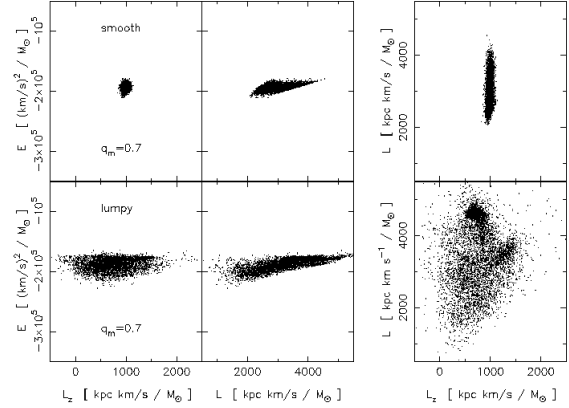}}{}
\caption{As    Figure~4,   but   in    a   flattened    halo   with
$q_m=0.7$.  Evidently, even  in  a flattened  halo,  it is  relatively
straightforward to distinguish the effect of the sub-clumps.}
\end{figure}

\begin{figure}
\ifthenelse{\UseFigs=1}{
\includegraphics[angle=0,width=\hsize]{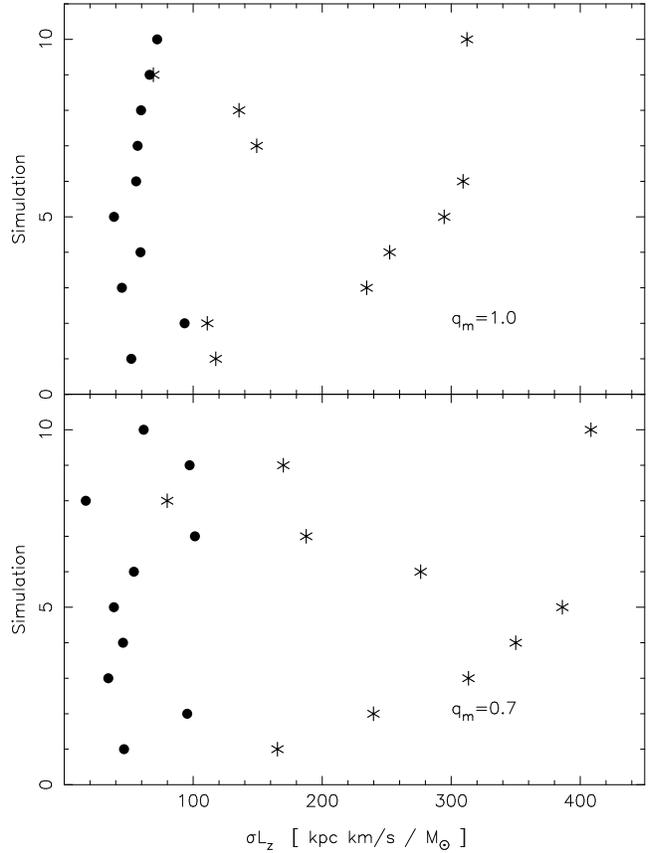}}{}
\caption{The dispersion in the z-component of angular momentum per unit mass
($\sigma L_z$) in several simulations. Filled circles denote the smooth halo
simulations, while star-symbols show the results with NFW sub-halos. The top
panel is for those simulations with a spherical mass distribution, while the
bottom panel  shows the results in  a flattened halo  $q_m=0.7$. Clearly, if
the Galactic halo has  substantial substructure as predicted by cosmological
simulations,  globular cluster  streams  should have  a  wide dispersion  in
$L_z$.}
\end{figure}

\begin{figure}
\ifthenelse{\UseFigs=1}{
\includegraphics[angle=270,width=\hsize]{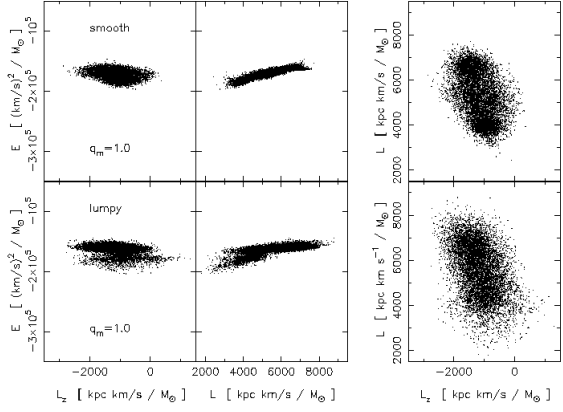}}{}
\caption{As Figure~4, but for the case of a stream which has already been
detected: that of the Sagittarius dwarf galaxy. Here, the difference between
the smooth and lumpy halos is much less marked due to the fact that the
initial model is much more massive, and has naturally a wide spread in $E$,
$L_z$ and $L$.}
\end{figure}

\end{document}